\documentclass[]{spie}  
\usepackage[]{graphicx,epsfig}

\title{A Fokker--Planck description for Parrondo's games}

\author{R. Toral\supit{a}, P. Amengual\supit{a} and S. Mangioni\supit{b}
\skiplinehalf
\supit{a}Instituto Mediterr\'aneo de Estudios Avanzados (IMEDEA) CSIC-UIB,\\ Ed. Mateu Orfila, Campus UIB, 07122-Palma de Mallorca, Spain\\
\supit{b}Departamento de F{\'\i}sica, Facultad de Ciencias Exactas y Naturales,\\ Universidad Nacional de Mar del Plata, De\'an Funes 3350, 7600 Mar del Plata, Argentina\\
}
\authorinfo{Further author information: Send correspondence to R. Toral, e-mail: raul@imedea.uib.es, http://www.imedea.uib.es}

\begin{document}
\maketitle
\begin{abstract}
We discuss in detail two recently proposed relations between the Parrondo's games and the Fokker--Planck equation describing the flashing ratchet as the overdamped motion of a particle in a potential landscape. In both cases it is possible to relate exactly the probabilities of the games to the potential in which the overdamped particle moves. We will discuss under which conditions current-less potentials correspond to fair games and vice versa. 
\end{abstract}

\section{Introduction} The Parrondo's Paradox\cite{HA99,HA99b,HA02} is a
combination of games inspired by the mechanism of Brownian ratchets\cite{R02}
and, more specifically, by the flashing ratchet\cite{AB94,PCPA94} which shows
that it is possible to use fluctuations (in the form of noise) to obtain
directed motion in the absence of any systematic macroscopic forces -or gradients-. The
physical picture of a flashing ratchet consists of a Brownian particle  under
the influence of a potential that is turned 'on' and 'off' either periodically
or stochastically.  Parrondo's paradox ``translates" this physical mechanism by
considering very simple losing -or fair- gambling games, whose alternation
results in a winning game. In the original, and still the simplest version, one
considers a player which tosses biased coins, such that one unit of ``capital"
is won (lost) if heads (tails) show up. Two games are combined. The first game,
game A, is such that there is the same probability of winning independently of
the capital of the player: $p_i=\frac{1}{2}\,,\forall i\,$, where $i$ denotes
the actual capital of the player. The second game, game B, has probabilities
which depend on whether the capital $i$ of the  player is or not a
multiple of $3$,
\begin{equation}\label{1}
p_i=\left\{ \begin{array}{cc}
	       \frac{1}{10} & \textrm{if}\,\, mod(i,3)=0\\
	       \\
	       \frac{3}{4} & \textrm{if}\,\, mod(i,3)\neq 0
 	      \end{array}
       \right.
\end{equation}
It is easy to show that both games $A$ and $B$ are fair games (``Parrondo" games), while the alternation (either random or periodic) of both games produces a winning game. Many other extensions of the games have been proposed, including a version with probabilities depending on the outcome of the previous games\cite{PHA00}\,, collective players\cite{T01,T03}\,, quantum games\cite{FNA02}\,, Ising systems\cite{M00}\,, pattern formation\cite{BLP02}\,.

Although, as stated before, Parrondo's games were \textit{inspired} by the
flashing ratchet, there was not a direct and precise connection between both, but only qualitative arguments. It has been only very recently that the
work by Allison and Abbott\cite{AA02} has established a quantitative relation
between the physical parameters (potential) of the flashing ratchet and the
probabilities of Parrondo's games, by discretizing conveniently the Fokker--Planck equation for the flashing ratchet and comparing it afterwards with the master equation for the games. An alternative relation has been put forward by us\cite{TAM03} using a somewhat different approach in which we identify the current directly in the master equation and then compare it with an {\sl ad hoc} discretization of the Fokker-Planck equation. In both cases it is possible to relate precisely the probabilities defining the games with the values of a discretized version of the physical potential that represents the ratchet.

In this paper we will study in detail the results given by the two proposed
relations, references\cite{AA02,TAM03}\,, between the games and the ratchets and we will give explicit results concerning these relations and, in particular, the finding of a suitable potential for the description of the games. The paper is organized as follows: in the next section we review the proposed relations between the games and the ratchets and write down the equivalence between the physical parameters of a ratchet and the games probabilities. We work in both directions: first, given the probabilities of the Parrondo's games, we obtain the Brownian particle potential; second, from a given potential, we extract the corresponding probabilities. In section \ref{III} we give specific examples using the original version of the games and a widely used ratchet potential. Some technical problems appearing for an even number of discretization points in our formalism are discussed in section \ref{IV}. Finally, section \ref{V} briefly summarizes the main results.

\section{Relation between the Fokker--Planck and Master equations} 
\label{II}
The first connection is that proposed by Allison and Abbott\cite{AA02} which we now review very briefly. These authors start by considering the following general Fokker--Planck equation\cite{HL84} for the probability $P(z,t)$ of a Brownian particle moving in a time-dependent one-dimensional potential $V(z,t)$
\begin{equation} \label{20}
D \frac{\partial^2P}{\partial z^2}-P \frac{\partial \alpha}{\partial z}-\alpha\frac{\partial P}{\partial
z}-\frac{\partial P}{\partial t}= 0,
\end{equation}
here $\alpha$ and $D$ are referred to as the infinitesimal first and second moments of diffusion, respectively; $D$ is 
considered to take a constant value --``Fick's law constant"-- $D\approx1.3 \times 10^{-9}\,m^2s^{-1}$ while $\alpha(z,t)$ is a function related to the applied potential $V(z,t)$  by 
\begin{equation} \label{21}
\alpha(z,t)=-u\frac{\partial}{\partial z}V(z,t),
\end{equation}
$u$ denotes the mobility of the Brownian particle given by 
\begin{equation}
u=\frac{Z_e}{6 \pi \eta a}\approx 51.9\times 10^{-9}m^2s^{-1} 
\end{equation} 
where $Z_e$ accounts for the electrical charge on the 
particle, $\eta$ is the kinematic viscosity of the solvent and $a$ is the effective radius of the particle.
The next step is to discretize Eq. (\ref{20}) using a finite difference approximation to obtain:
\begin{equation}
P_{i,j}=a_{-1}^{i,j}\cdot P_{i-1,j-1}+a_{0}^{i,j}\cdot P_{i,j-1}+a_{+1}^{i,j}\cdot P_{i+1,j-1} \label{22}
\end{equation}
where
\begin{equation}
a_{-1}^{i,j}=\frac{\frac{D\tau}{\lambda^2}+\frac{\alpha(i,j)
\tau}{2\lambda}}{\frac{\alpha(i+1,j-1)-\alpha(i-1,j-1)}{2\lambda}\tau+1} \label{23}
\end{equation}
\begin{equation}
a_{0}^{i,j}=\frac{-2\frac{D\tau}{\lambda^2}+1}{\frac{\alpha(i+1,j-1)-\alpha(i-1,j-1)} 
{2\lambda}\tau+1} \label{24}
\end{equation}
\begin{equation}
a_{+1}^{i,j}=\frac{\frac{D\tau}{\lambda^2}-\frac{\alpha(i,j)
\tau}{2\lambda}}{\frac{\alpha(i+1,j-1)-\alpha(i-1,j-1)}{2\lambda}\tau+1} \label{25}
\end{equation}
Index $i$ denotes the discretized space, $z=i\lambda$, whereas $j$ denotes the discretized time, $t=j\tau$, being $\lambda$ and $\tau$, respectively, the space and time discretization steps. Therefore $P_{i,j}$ stands for $P(i\lambda,j\tau)$, $\alpha(i,j)$ stands for $\alpha(i\lambda,j\tau)$ and similar relations between the continuum and discretized version of other functions.

This discretized form (\ref{22}) of the Fokker--Planck equation is compared to a general master equation for any of the gambling games used in the Parrondo's paradox. This general master equation is 
\begin{equation}
P_{i,j}=p_{i-1}\cdot P_{i-1,j-1}+q_i\cdot P_{i,j-1}+(1-p_{i+1}-q_{i+1})\cdot P_{i+1,j-1} \label{26}
\end{equation}
where $P_{i,j}$ denotes the probability that the player has a capital $i$ at
the $j$ tossing of the coin. In this way, the time $t$ of the Brownian particle
becomes the number $j$ of tossed coins and the position $z$ becomes the value
of the capital $i$. In this general master equation $p_i$ is the probability of winning given that the value of the capital is $i$, and $q_i$ is the
probability of remaining the same (neither losing nor winning) given that the
value of the capital is $i$. In accordance with the rules of the Parrondo games we will set $q_i=0$ since at each tossing of the coins the capital increases or decreases, but never remains the same. Notice that, in accordance again with the rules of the Parrondo's games, we are assuming that the probabilities $p_i$ do not depend on time (the index $j$). This implies that the discretized version of the first moment of diffusion $\alpha(z,t)$ does not depend on the index $j$ either, $\alpha(z,t)\to \alpha_i$. Moreover, the probabilities must satisfy $p_{i+L}=p_i$ for a given $L$ ($L=3$ in the original version of the games).

Combining equation (\ref{22}) and equation (\ref{26}) we get 
\begin{equation}
\frac{p_{i-1}}{1-p_{i+1}}=\frac{a_{-1}}{a_{+1}}=\frac{1+\frac{\lambda}{2D\tau}\alpha_{i}}{1-\frac{\lambda}{2D\tau}\alpha_{i}}
\end{equation}
from where it can be obtained an expression for $\alpha_{i}$
\begin{equation}
\label{alfa}
\alpha_{i}=\frac{2D}{\lambda}\,\frac{p_{i-1}-(1-p_{i+1})}{p_{i-1}+(1-p_{i+1})}.
\end{equation}
Finally, the discretized values of the potential are obtained by using a convenient discretization of Eq. (\ref{21}), namely:
\begin{equation}
V_i=-\frac{2D}{u}\sum^i_{k=0}\frac{1-(\frac{1-p_{k+1}}{p_{k-1}})}
{1+(\frac{1-p_{k+1}}{p_{k-1}})}\label{2}
\end{equation}
This fundamental relation allows one to obtain the discretized version of the physical potential $V_i$ in terms of the probabilities $p_i$ of the games. It is possible to invert it to find the $p_i$ in terms of the $V_i$. If one wants to recover a Parrondo-type game, the inverse equations need to be solved under the condition that $p_{i+L}=p_i$ for a suitable value of $L$.
For the case of L=3  we obtain the following expressions for the $p_i$: 
\begin{eqnarray}
p_0 &= &-\frac{(-1+V'_0-V'_1)(-1+V'_0+V'_1+3V'_0V'_1-V'_2-3V'_0V'_2)}{2(-1+{V'}_0^2-V'_0V'_1+{V'}_1^2-V'_1V'_2)}\nonumber\\
p_1 &= &-\frac{(-1+3{V'}_0^2+V'_1-V'_0(2+3V'_1))(-1+V'_1-V'_2)}{2(-1+{V'}_0^2-V'_0V'_1+{V'}_1^2-V'_1V'_2)}\nonumber\\
p_2 &= &-\frac{(1+V'_0)(1-V'_0+2V'_1+3V'_0V'_1-3{V'}_1^2-V'_2-3V'_0V'_2+3V'_1V'_2)}{2(-1+{V'}_0^2-V'_0V'_1+{V'}_1^2-V'_1V'_2)}\label{13}
\end{eqnarray}
\vspace{0.3 cm}
where $V'_i=-\frac{u}{2D}V_i$ .\\
A second connection between the Fokker--Planck equation and the master equation is given by our treatment in reference\cite{TAM03}\, (see also reference\cite{HKK02}). Our approach to this problem is, in some sense, complementary to that of reference\cite{AA02} since our starting point is not the Fokker--Planck equation but rather the master equation (\ref{26}) which we write in the form:
\begin{equation}
P_{i,j}=a_{-1}^i P_{i-1,j-1}+a_0^iP_{i,j-1}+a_1^iP_{i+1,j-1} \label{27}
\end{equation} 
where $a_{-1}^i$ is the probability of winning when the capital is $i-1$, $a_{1}^i$ is the probability of losing when the capital is $i+1$, and $a_0^i$ is the probability that the capital $i$ remains unchanged (a possibility not considered in the original Parrondo games). To ensure the conservation of probability ($\sum_iP_{i,j}=\sum_i P_{i,j-1}$) the coefficients $a_i$ must satisfy  
\begin{equation}
a_{-1}^{i+1}+a_0^i+a_1^{i-1}=1 \label{28}
\end{equation}
Replacing this equation in (\ref{27}) it is possible to write the master equation in the form of a continuity equation for the probability 
\begin{equation}
P_{i,j}-P_{i,j-1}=-\left[J_{i+1,j}-J_{i,j}\right] \label{29}
\end{equation}
where the current $J_{i,j}$ is given by:
\begin{equation}
\label{current}
J_{i,j}=\frac{1}{2}\left[F_i P_{i,j}+F_{i-1}P_{i-1,j}\right]-\left[D_iP_{i,j}-D_{i-1}P_{i-1,j}\right] \label{30}
\end{equation}
and
\begin{equation}
F_i = a_{-1}^{i+1}-a_1^{i-1},\hspace{2.0cm} 
D_i = \frac{1}{2}(a_{-1}^{i+1}+a_1^{i-1})  \label{31}
\end{equation}
We can relate the previous coefficients (\ref{31}) with their analogous terms corresponding to a discretization of the Fokker--Planck equation for a probability $P(x,t)$
\begin{equation}
\frac{\partial P(x,t)}{\partial t}=-\frac{\partial J(x,t)}{\partial x} \label{32}
\end{equation}
with a current
\begin{equation}
J(x,t)=F(x)P(x,t)-\frac{\partial [D(x)P(x,t)]}{\partial x} \label {33}
\end{equation}
with general drift, $F(x)$, and diffusion, $D(x)$. If $\tau$ and $\lambda$ are, respectively, the time and space discretization steps, such that $x=i\lambda$ and $t=j\tau$, it is clear the identification
\begin{equation}
F_i \longleftrightarrow \frac{\tau}{\lambda} F(i\lambda), \hspace{1.0truecm}
D_i \longleftrightarrow \frac{\tau}{\lambda^2} D(i\lambda) \label{34}
\end{equation}
Considering the case $a^i_0=0$ and since $p_i=a_{-1}^{i+1}$ we have 
\begin{equation}
D_i\equiv D=1/2\hspace{2.0truecm} F_i=-1+2p_i \label{35}
\end{equation}
and the current takes the following form  : $J_{i,j}=-(1-p_i)P_{i,j}+p_{i-1}P_{i-1,j}$ . The latter expression is nothing but the probability flux from site $i-1$ to site $i$.
We now define a potential $V_i$ in terms of $F_i$, or, equivalently, $p_i$ 
\begin{equation}
\label{3}
V_i=-\frac{1}{2}\sum_{k=1}^i\ln\left[\frac{1+F_{k-1}}{1-F_k}\right]=-\frac{1}{2}\sum_{k=1}^i\ln\left[\frac{p_{k-1}}{1-p_k}\right] \label{36}
\end{equation}
we have adopted the, otherwise arbitrary, value $V_0=0$. This equation is our main result concerning the relation between the games probabilities $p_i$ and the discretized version of the potential $V_i$. In the next section we will compare both expressions (\ref{2}) and (\ref{3}).

We can obtain the stationary probability distribution $P_i^{st}$ through the recurrence relation derived from equation (\ref{30}) using a constant current $J_i=J$, together with the boundary condition $P_i^{st}=P_{i+L}^{st}$. The result is:
\begin{equation}
P_i^{st}=N {\rm e}^{-2V_i}\left[1- \frac{2 J}{N} \sum_{j=1}^i \frac{{\rm e}^{2V_j}}{1-F_j}\right] \label{37}
\end{equation}
and a current
\begin{equation}
\label{38}
J=N\frac{{\rm e}^{-2V_L}-1}{2\sum_{j=1}^L \frac{{\rm e}^{2V_j}}{1-F_j}}
\end{equation}
$N$ is nothing but a normalization constant obtained with $\sum_{i=0}^{L-1} P_i^{st}=1$. It is interesting to note that this expression for the current coincides in the case $L=3$ with the known formula\cite{LAAS02}:
\begin{equation}
\label{current3}
J=\frac{p_0p_1p_2-(1-p_0)(1-p_1)(1-p_2)}{2+p_0p_1p_2-(1-p_0)(1-p_1)(1-p_2))}
\end{equation}

The inverse problem of obtaining the game probabilities in terms of the potential requires solving Eq. (\ref{36}) with the boundary condition $F_0=F_{L}$. For that purpose we must solve the following set of equations: 
\begin{equation}
F_{i+1}=1-{\rm e}^{\Delta_i}-F_i{\rm e}^{\Delta_i}  \label{39a}
\end{equation}
where $\Delta_i=V_{i+1}-V_i$. We recall first that the general solution of the set of recurrence relations:
\begin{equation}
x_i=a_i+b_i\cdot x_{i-1} \hspace{2.0cm} i=0,1,\dots\label{40a} 
\end{equation} 
with initial condition $x_0$ is 
\begin{equation}
x_n=\big [\prod_{k=1}^n b_k \big ]\cdot x_0+\sum_{j=1}^n a_j\big[\prod_{k=j+1}^n b_k\big]
\end{equation}
Identifying in Eq. (\ref{39a}):
\begin{eqnarray}
a_i& =& 1-e^{\Delta_{i-1}}=1-e^{V_i-V_{i-1}},\nonumber \\
b_i& =& -e^{\Delta_{i-1}}=-e^{V_i-V_{i-1}}
\label{40}
\end{eqnarray}
and 
\begin{eqnarray}
&\prod_{k=1}^n b_k &= (-1)^{n-j} e^{V_n-V_j},\nonumber \\
&\prod_{k=j+1}^n b_k &= (-1)^n e^{V_n-V_0}
\label{41}
\end{eqnarray}
we arrive at the following expression
\begin{equation}
\label{39}
F_i=(-1)^i{\rm e}^{2V_i}\left[\frac{\sum_{j=1}^{L}(-1)^j[{\rm e}^{-2V_j}-{\rm e}^{-2V_{j-1}}]}{(-1)^L{\rm e}^{2(V_0-V_L)}-1}+\sum_{j=1}^{i}(-1)^j[{\rm e}^{-2V_j}-{\rm e}^{-2V_{j-1}}]\right]
\end{equation}
which, via $p_i=(1+F_i)/2$, allows one to obtain the probabilities $p_i$ in terms of the potential $V_i$. It is clear that the additional condition $p_i\in[0,1],\,\forall i$ must be fulfilled by any acceptable solution.

Summing up, given a set of probabilities $(p_0,\dots,p_{L-1})$ defining a Parrondo game, there are two possibilities for the stochastic potential $V_i$, using either Eq. (\ref{2}) or Eq.(\ref{3}).  From any one of these, the inverse problem can be solved in order to find the probabilities given a stochastic potential. 
In the next section we give explicit examples of these procedures.

\section{Examples}
\label{III}

\begin{figure}[!htb]
\centerline{
\includegraphics[width=6cm,height=10cm,angle=-90]{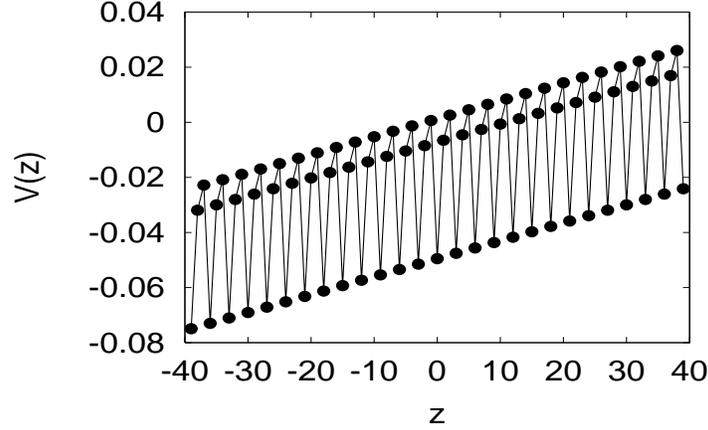}}
\caption{\label{fig1} Potential $V_i$ derived from the original Parrondo game $B$, Eq. (\ref{1}) using the relation between the potential and the probabilities given by Eq.(\ref{2}).}  
\end{figure}

We use the game $B$ defined by Eq.(\ref{1}). This is equivalent to take
$L=3$ and $p_0=1/10,\,p_1=p_2=3/4$ supplemented by $p_{i+3}=p_i$. The
stochastic potential derived using (\ref{2}) is plotted in figure \ref{fig1}.
It is clear that this potential is not periodic. Therefore, it is the
discretized version of a nonperiodic continuum potential $V(z)$. The potential
tilt, in the continuum case, will induce a non--zero current. This is in
contrast with the condition of fairness for game $B$ which implies that there
is no net gain when playing that game. Instead, this kind of potential would
be characteristic of an unfair game with a current J to the left, i.e.
a losing game.

If we use Eq.({\ref{3}) instead we obtain the following values for the discretized potential $V_0=0,\,V_1=\ln\big(\frac{5}{2}\big),\,V_2=\ln\big(\frac{5}{6}\big)$ together with the periodicity condition $V_{i+3}=V_i$. Notice that this periodicity condition is equivalent to 
\begin{equation}
 p_0 p_1 p_2=(1-p_0)(1-p_1)(1-p_2)
\label{5}
\end{equation}
which is the condition of B being a fair game. In fact, the general condition for a fair game can be obtained through the discrete time Markov chain analysis as\cite{HA02}  
\begin{equation} 
\prod^{L-1}_{i=0} p_i=\prod^{L-1}_{i=0} (1-p_i).
\label{10}
\end{equation} 
which is automatically fulfilled by (\ref{36}) for periodic potentials $V_i=V_{i+L}$.

The resulting values of this potential are plotted in figure \ref{fig2}. In this case, we note that the potential can be thought of as the discretized version of a null current continuous potential. 

\begin{figure}[!htb]
\centerline{
\includegraphics[width=6cm,height=10cm,angle=-90]{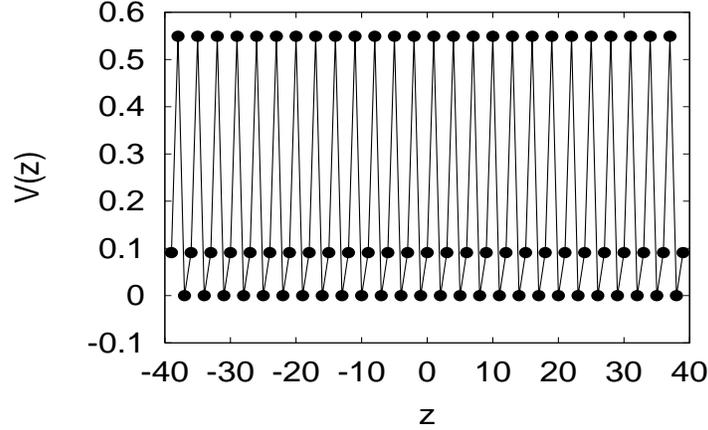}}
\caption{\label{fig2}Similar to figure \ref{fig1} using instead relation between the potential and the probabilities given by Eq.(\ref{3}).}
\end{figure}

We now turn to the inverse problem: given the
discretized version of the potential , $V_i$\,, obtain the corresponding probabilities $p_i$ using Eqs. (\ref{13}) or (\ref{39}). We consider the widely used ratchet potential\cite{R02} $V(x)$: 
\begin{equation}
V(x)=A\,\left[\sin\left(2\,\pi\,x\right)+\frac{1}{4}\sin\left(4\,\pi\,x\right)\right]
\label{7}
\end{equation} 
with an amplitude $A=0.01$. We discretize it by using $\lambda=1/3$ such that there are $L=3$ points per period and $V_i=V(i\lambda)$.We obtain the following sets of probabilities.Using equation (\ref{13}): 
\begin{equation}
\begin{array}{rcl}
p_0& =&0.567082\\
p_1& =&0.736037\\
p_2& =&0.263963.
\end{array}
	\label{8}
\end{equation}
Using (\ref{current3}) these values yield a current $J=0.0118792$.\\
Using equation (\ref{39}) we obtain instead 
\begin{equation}
\begin{array}{rcl}
p_0& =&0.493504\\
p_1& =&0.503279\\
p_2& =&0.503216.
\end{array}
	\label{9}
\end{equation}
In this case the current is $J=0$ as it corresponds to a fair game.

\section{The case of $L$ even}
\label{IV}

A problem arises when finding the probabilities $p_i$ using (\ref{39}) for a periodic potential (corresponding to a fair game) when the number of points $L$ is even. This is obvious since the periodicity condition $V_L=V_0$ gives a zero value for the denominator $(-1)^L {\rm e}^{2(V_0-V_L)}-1$. In order to be able to find solutions for the probabilities, the numerator has to vanish as well. This is equivalent to the condition:
\begin{equation}
\sum_k e^{-2V_{2k}}=\sum_k e^{-2V_{2k+1}}
\label{11}
\end{equation}
which, in terms of the stationary probabilities, becomes:
\begin{equation}
\sum_k P^{st}_{2k}=\sum_k P^{st}_{2k+1}.
\label{12}
\end{equation}
This condition implies that one can have a fair game in the case of an even number $L$ only if the  probability of finding an even value for the capital  equals that of finding an odd value. To our knowledge, this curious property, which emerges naturally from the relation between the potential and the probabilities, has not been reported previously.

It turns out that one has to be careful when discretizing a periodic potential $V(x)$ in order to preserve this property. Otherwise, there will be no equivalent Parrondo game with zero current. The simple identification $V_i=V(i\lambda)$ might not satisfy this requirement, but we have found that a possible solution is to shift the origin of the $x$-axis, i.e. setting $V_i=V((i+\delta)\lambda)$ for a suitable value of $\delta$. For example, in Fig.\ref{delta} we plot the difference
\begin{equation}\label{eqd}
d(\delta)=\sum_i e^{-2V((2i+\delta)\lambda)}-\sum_i e^{-2V((2i+1+\delta)\lambda)}\end{equation}
 as a function  of $\delta$ in the case of the potential (\ref{7}) and $\lambda=1/4$ (which corresponds to $L=4$ points per period). We see that there is  only one value that accomplishes $d(\delta)=0$, namely $\delta=-0.068616$.

\begin{figure}[!htb]
\centerline{\includegraphics[width=6cm,height=10cm,angle=-90]{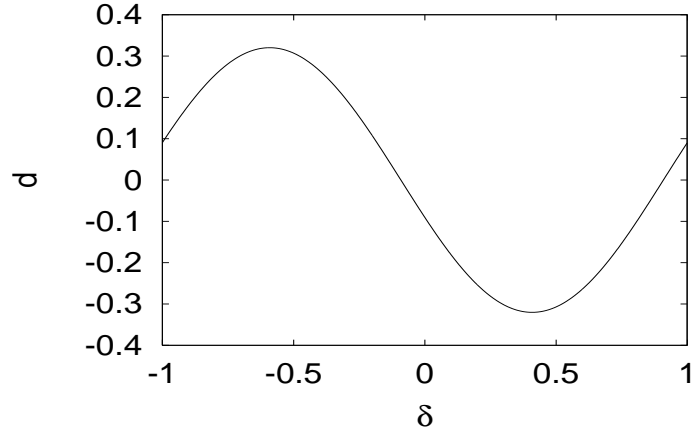}}
\caption{\label{delta}Plot of $d(\delta)$ as given by Eq. (\ref{eqd}) versus displacement $\delta$. The unique zero crossing is at $\delta=-0.068616$.}
\end{figure}

Once the proper value of $\delta$ is found, it follows from Eq. (\ref{39}) that there are infinitely many solutions for the probabilities. They can be found by varying, say, $p_0$, such that for each value of $p_0$ we will get a set of probabilities $p_i$. Solutions satisfying the additional requirement that $p_i\in[0,1],\,\forall i$, will exist only for a certain range of values of $p_0\in[0.0025,0.68]$. Some of the different solutions are plotted in figure \ref{2a}. Some numerical values are :
\begin{itemize}
\item $p_0=0.125$, $p_1=0.8167766$, $p_2=0.3927740$, $p_3=0.7082539$  
\item $p_0=0.25$, $p_1=0.6335531$, $p_2=0.5289900$, $p_3=0.6070749$
\item $p_0=0.3525$, $p_1=0.4833099$, $p_2=0.6406871$, $p_3=0.5241081$
\item $p_0=0.50$, $p_1=0.2671062$, $p_2=0.8014221$, $p_3=0.4047168$
\end{itemize}

\begin{figure}[!htb]
\centerline{\includegraphics[width=6cm,height=10cm,angle=-90]{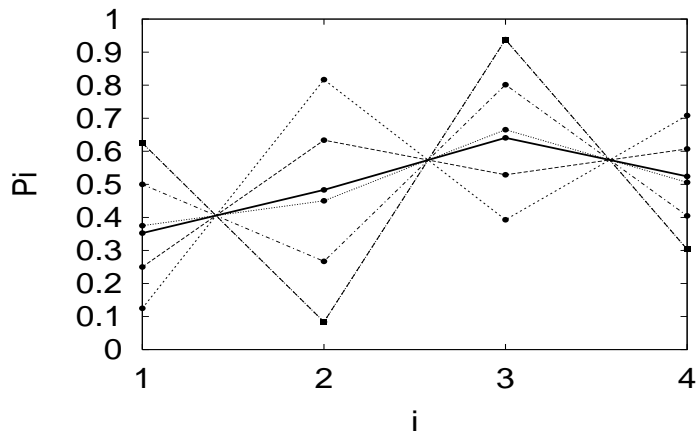}}
\caption{\label{2a}Multiple solutions for the probabilities $p_i$ obtained with equation (\ref{3}) for a potential like (\ref{7}) with $A=0.3$\,, $\lambda=\frac{1}{4}$\,, $\delta=-0.068616$ varying the value of $p_0$. The continuous line corresponds to the ``optimal" solution, $p_0=0.3525$ (see the text).}
\end{figure}

An additional criterion to chose between the different sets of probabilities is to impose the maximum ``smoothness" in the distrbution of the $p_i$'s. For instance, one could minimized the sum $\sum_{i=0}^{L-1} (p_{i+1}-p_i)^2$. In our example this criterion yields $p_0=0.3525$ and the other values follow from the previous table.}

\section{Conclusions}
\label{V}
In summary, we have compared the results given by two recently proposed relations between the master equation describing a Parrondo game and the Fokker--Planck equation for a ratchet. While the approach of reference [13] discretizes first the Fokker--Planck equation and then relates it to a master equation, our own approach takes as a starting point the master equation and then finds the {\sl ad hoc} (although consistent) discretization of the Fokker--Planck equation that yields exactly the same master equation. We have found for the two proposals the potential that corresponds to a particular set of the game probabilities, as well as the probabilities that correspond to a a widely used ratchet potential. We have analyzed in detail the case of an even number of points in the discretization obtaining that for a given potential there are many possible probabilities compatible with it. The discretization, in this case, has to fulfill the property that the stationary probability of having an even value of the capital is equal to that of having an odd value. 

{\noindent \bf Acknowledgements:}  This work is supported by the Ministerio de
Ciencia y Tecnolog{\'\i}a (Spain) and FEDER, projects BFM2001-0341-C02-01 and
BFM2000-1108.

\end{document}